\documentclass[10pt,conference]{IEEEtran}

\usepackage{amsmath,amssymb,amsfonts}
\usepackage{algorithmic}
\usepackage{cite}
\usepackage{graphicx}
\usepackage{textcomp}
\usepackage{xcolor}

\def\BibTeX{{\rm B\kern-.05em{\sc i\kern-.025em b}\kern-.08em
    T\kern-.1667em\lower.7ex\hbox{E}\kern-.125emX}}

\newcommand{\orr}{$\mbox{OR}(r)$\hspace{1pt}}

\begin{document}

\title{Orthodromic Routing and Forwarding for Large Satellite Constellations}

\author{\IEEEauthorblockN{Peter Ashwood-Smith\IEEEauthorrefmark{1},
Bill McCormick\IEEEauthorrefmark{2}}
\IEEEauthorblockA{Huawei Technologies Canada \\
Email: \IEEEauthorrefmark{1}peterashwoodsmith@icloud.com,
\IEEEauthorrefmark{2}bill.mccormick@huawei.com}}
\maketitle

\begin{abstract}
Low earth orbit satellite constellations with inter-satellite links (ISLs) are currently being developed and deployed.
The availability of ISLs provides the capability to route across the satellite constellation, rather than using the satellite as a 
single hop in a bent-pipe configuration.    We present a fully distributed solution to routing and forwarding which we call 
Orthodromic Routing (\orr). \orr routing is built on a foundation of both geographic and link state routing to create a hybrid protocol which 
scales to enormous constellations with excellent failure handling.   Our work includes an addressing and forwarding plane for \orr  
which can be implemented in hardware in a highly parallel manner to achieve line rates while only requiring a bounded number of 
forwarding table entries.
\end{abstract}

\section{Introduction}

Large scale low earth orbit satellite constellations are being introduced to extend communications to rural and low coverage areas
worldwide.   This paper proposes orthodromic routing, an inter-satellite routing mechanism using techniques from link state routing protocols and making use of the geometric properties of the satellite constellation.

When two points on the earth are covered by the same satellite, or if the satellites are unable to communicate with each other, inter-satellite 
routing is not required. LEO and MEO satellite constellations, to achieve their full potential, require ISLs which, because of the relative motion of the satellites also require steering and precise aiming capabilities. This steering can be achieved with phased array antennas for RF based ISLs or mechanical means for free space optics based ISLs \cite{isl}.  

Sufficient coverage and throughput requires thousands of satellites in a constellation. Modern routing protocols such as  
OSPF \cite{ospf} and IS-IS \cite{isis} require hierarchy to  scale to these numbers but hierarchy is based on fixed membership, 
fixed borders, and hierarchical forwarding. For example, flat link-state networks are typically recommended to be under a few 
hundred routers with a thousand being possible but relatively uncommon.

\begin{figure}
\centering
\includegraphics[width=0.9\linewidth]{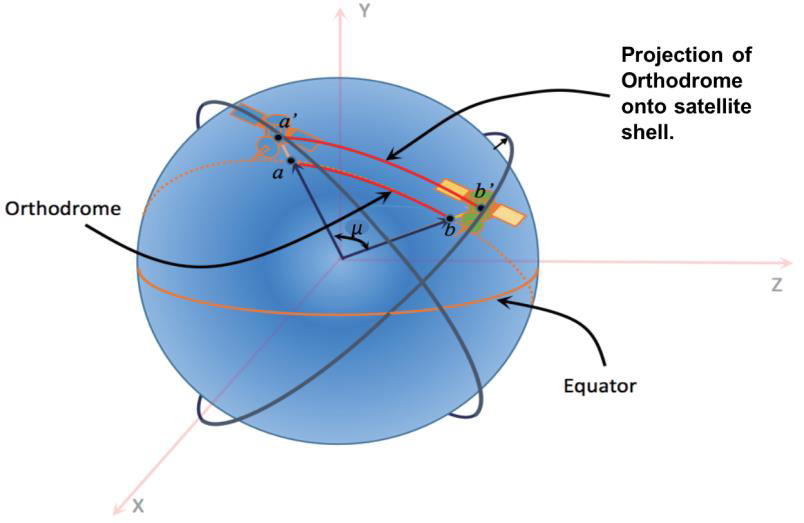}
\caption{An orthodrome is a portion of a great circle at the surface of the earth, however we use the term independently of the radius of the sphere on which we are routing to conveniently express the shortest possible path on that sphere’s surface between two points.}
\label{fig:orthodrome}
\end{figure}

Different constellations are possible but most are inclined circular orbits which span the 360 degrees around the equator.  
Figure \ref{fig:orthodrome} shows a relatively steep angle to the equator which increases density of coverage at the expense 
of the poles.   This is referred to as a Walker Delta constellation.

Different coverage requirements dictate the orbital angles and thus the degree of polar coverage. The density of satellites varies from 
sparse at the equator to high density at the latitudes corresponding to the chosen angle of inclination  \cite{coverage}. 

Such constellations with 4-way interconnected links (ISLs) pose enormous challenges to traditional hop by hop packet routing schemes since:
\begin{itemize}
\item The inter-satellite links (ISLs) go up and down when relative satellite motion is high such as near the poles and also when 
satellites in close proximity are travelling in almost opposite directions. This results in frequent link state changes requiring rapid flood 
rates \cite{scale1}.
\item Centralized routing schemes allow for distribution of state but the volume and frequency of changing state is 
impractical for a software defined networking (SDN) type solution unless a source routing mechanism is used to eliminate state distribution. 
Source routing requires that source routes be distributed to the terminals, or at least the first hop satellites, which is cumbersome and it 
creates a bootstrap problem as a functioning routing system is required to distribute the state that enables routing 
\cite{scale2}.
\end{itemize}

Constellations with multiple orbital shells at different elevations may have difficulty connecting between shells using
inter-satellite links due to the large relative velocity between satellites orbiting at different elevations.

It is also expected that due to the sheer size of the Internet routing tables, differing aggregation methods and the dynamics of the 
combined system that satellite networks must provide a well isolated private network over which the public Internet or other 
private networks (e.g. 3GPP back-haul) will tunnel and interwork.

In \cite{aodv}, the authors propose adapting the ad hoc on-demand distance vector routing protocol for use in 
satellite constellations and conclude it requires augmentation with a proactive protocol.  By contrast, \orr is a proactive 
link state routing protocol, and we restrict the use of flooding from global scope
 to local scope to improve control plane throughput.
 
Older satellite routing techniques rely on a virtual node approach \cite{virtual-node}, where 
the moving satellites are replaced by static, virtual nodes with fixed locations with respect to the earth.   
These consider smaller numbers of satellites, while \orr is designed for massive physical topologies.
In \cite{demand-island}, the authors update this approach  with demand islands, which provide a highly
 available method of connecting ground terminals to the satellite network over short distances but do not allow 
 arbitrarily long paths that can take advantage of the low delay possible via multihop ISL routing.
 
In \cite{link-state}, the authors propose a link state based protocol in smaller satellite networks which handles 
isolated link failures.  OR handles arbitrary link failure combinations and uses localized flooding to avoid message storms.
 
Centralized solutions to these challenges for relatively small constellations include ground based computations of end to end 
redundant pairs of scheduled source/segment routes which are then uploaded to the satellites and which may be invalidated by in 
orbit flood events for fast distributed failure recovery.

In this paper we design a scalable distributed routing system for thousands of satellites. The system needs to solve the following 
problems:
\begin{enumerate}
\item Floods need to be limited to considerably fewer nodes than the full constellation -- ideally a maximum of one thousand 
which is a reasonable upper bound on flood type protocols.
\item Addressing which allows selection of gateways at the edges of those flood regions even while the gateways are constantly 
changing and moving.
\item An address aggregation scheme that eliminates the need to store the private terminal ground based addresses in every 
satellite and certainly not global IP addresses.
\item A forwarding mechanism that can be implemented in hardware/software at line rates.
\end{enumerate}

Orthodromic Routing solves the above problems by trading some packet loss when there are sufficiently high failure rates in the 
ISL mesh against massive scalability. A family of algorithms \orr is defined where $r$ indicates the size of the flood 
radius in hops (and hence the size of ISL hole that can be circumnavigated). We then show how forwarding can be implemented 
in these \orr algorithms. We also show a simple proof of correctness and simulation results 
of the required $r$ radius choice vs the failure probability of ISLs.

Essentially \orr can be thought of as shortest path routing on the unit sphere. An orthodrome is the segment of the great circle (shortest surface path) between two points on the earth and $r$ indicates how many hops ahead the algorithm can look.  It can be helpful to understand \orr relative to existing well known routing protocols.

\begin{itemize}
\item $\mbox{OR}(1)$ is just geographic routing. Forward to closest neighbor within 1 hop.
\item $\mbox{OR}(r)$ behaves like link state when the destination is within r hops.
\item $\mbox{OR}(r)$ behaves like a multi-level routing protocol picking an appropriate gateway but with overlapping areas and dynamic gateways 
when the destination is not within r hops.
\item $\mbox{OR}(\infty)$ behaves like an impractically large link state protocol.
\end{itemize}

We now introduce the major components of OR in more detail.

\section{System Design}
\begin{figure}
\centering
\includegraphics[width=0.9\linewidth]{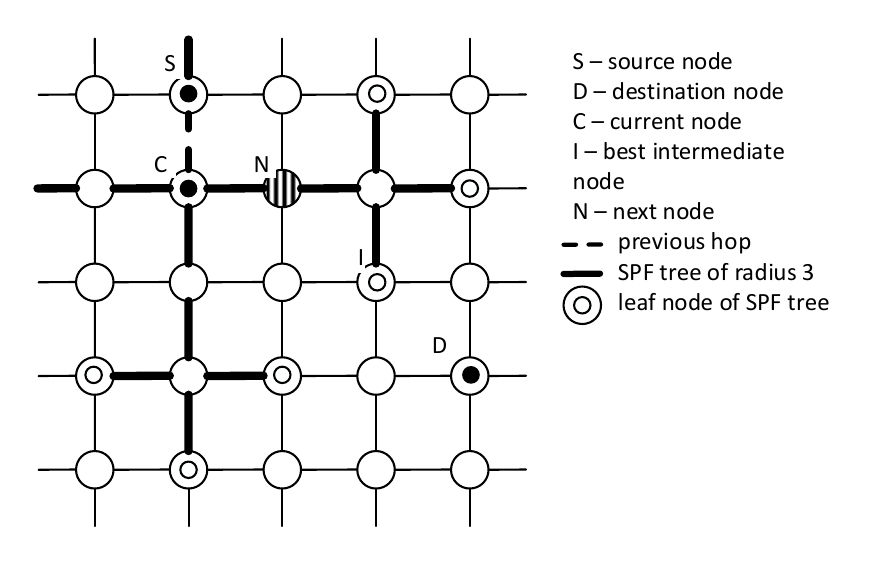}
\caption{The OR(r) algorithm simplified.  The SPF computations are not done per packet but are pre-computed.}
\label{fig:algorithm}
\end{figure}

The system model is shown in Figures \ref{fig:orthodrome} and \ref{fig:algorithm}.     Figure \ref{fig:orthodrome}
illustrates two orbits, and an orthdrome connecting a position on each orbit.
Figure \ref{fig:algorithm} shows part of the satellite mesh, with source and destination satellites.

To compute addresses, the satellite constellation is modelled as a unit sphere.   The address of a satellite is
constructed from a vector pointing to the satellite's location on the unit sphere.   The distance between two addresses is measured
as the angle $\mu$ between the two vectors as shown in Figure \ref{fig:orthodrome}.

The satellite network is modelled as a graph.   Without loss of generality,
we assume that nodes are of degree 4.   (Node degree will decrease in the presence of link failures).   When routing a packet through
the network, $S$ designates the source node of the packet, and $D$ designates the destination node of the packet.   $C$ is used to
denote the current node.  $I_k$ denotes an intermediate node which is also the leaf node of a shortest path first tree computed from
$C$.   The SPF tree has radius $r$ (in Figure \ref{fig:algorithm}, the radius is set to three).  $N$ denotes the next node after $C$ in the path.

\subsection{Addressing}

\orr source, destination and satellite addresses consist of two fields. The most important part, the location, is expressed 
as $\left<x, y, z\right>$ coordinates in $R^3$  lying on the surface of the unit sphere and can be encoded into
 an IP4 or IP6 address in different manners. The  precision of an address must be sufficient to differentiate 
 between two satellites in close proximity. 

The second portion of the address uniquely identifies the satellite or terrestrial device.   Each destination, D is 
reachable at some location $\left<x, y, z\right>$.   In addition it has a non-aggregatable identifier (similar to a MAC address)
that uniquely identifies the device.  A complete OR address has the form
\[
\left[ \mbox{ID} | \left<x,y,z \right> \right]
\]
\orr uses the location to get to a satellite close to the destination, and, once there, it uses the ID to identify the specific 
destination at that satellite.

An IP in IP encapsulation could be used where the outer addressing is private and of the form given above with respect to the 
satellite operator and the inner IP addressing is the private or public IP address of the traffic using the satellite constellation 
as a  tunnel.

OR uses a simplified metric, $\hat{\mu},$ for the distance between two addresses A and B.   Start with the angle between their
locations $\mu$ as a proxy for the distance on the surface of the unit sphere (Figure \ref{fig:orthodrome}). 
This requires the assumption that all satellites are at the same altitude.  Should multiple layers of satellites be used additional layers 
of routing and addressing are required to move between the shells, however each shell can still function
using OR as described in this paper.

The simplified metric, $\hat{\mu}$ is derived from the angle as follows.   Let $A$ and $B$ be two unit vectors.
\begin{eqnarray*}
A \cdot B &=& \lvert A \rvert \lvert B \rvert \cos(\mu) \\
\mu &=& \cos^{-1}( A \cdot B)
\end{eqnarray*}
$\mu$ could be used as a distance metric for routing.    The distance metric will be used in high performance forwarding operations,
so it's desirable to make the metric as computationally simple as possible, while still preserving relative distances.    
Making use of the fact that $\cos^{-1}()$ is monotonic increasing on $[-1, 1]$,
 approximate $\cos^{-1}()$ as a linear function
\[
\cos^{-1}(A \cdot B) \approx -\frac{\pi}{2} A \cdot B + \frac{\pi}{2}
\]
The coefficient $-\pi/2$ and the offset $\pi/2$ are not necessary to preserve relative distances, so they can be removed from
the final metric giving
\[
\hat{\mu} = - A \cdot B
\]
This metric is computationally simple to implement, requiring 3 multiplications and 2 additions using a componentized dot product:
\begin{equation}
-(A_x B_x + A_y B_y + A_z C_z )
\end{equation}

In OR(r), address aggregation occurs when addresses are close to each other with respect to this distance 
metric $\hat{\mu}$ (implying they are also close with respect to $\mu$, the angle between them).

Ground device addresses do not change for extended periods of time, as ground stations are either stationary, or
moving slowly relative to satellies and  
they aggregate into the coarser satellite’s addresses, negating the need for every satellite to store every 
ground reachable address. It's sufficient for a satellite to know its own address 
and those within its radius to be able to forward to any other satellite or ground station no matter the size
of the constellation.  Once a packet arrives close to the destination, forwarding based on the device
ID can be done since attached devices can be known within a very small radius of the attaching satellite at very low 
messaging overhead.   This implies when a device attaches to a satellite ideally it should be known by at least the 4 
neighboring satellites so that different routes via those 4 penultimate satellites will function correctly.

\subsection{Forwarding}

If a satellite determines that the ID of the destination is locally reachable (currently attached to this satellite) it does not route the packet to another satellite and instead local forwarding must deliver the packet down to the ground terminal based on that ID. This local operation is out of scope for this paper as it is essentially current bent pipe forwarding.

If the ID is not locally known the packet must be forwarded to one of the 4 adjacent satellites via the ISL link.  The packet needs to
follow a path that eventually will get closer to the destination.  If no other satellite is closer the packet will be dropped.

A forwarding table, created by the \orr computations, contains two columns. The first column contains the current addresses of 
the $K$ satellites within $r$-hops of the local node.  The second column is the ISL link index (1-4) to follow to reach that satellite in the 
least number of hops.  This second column is easily derived from the SPF tree of depth $r$ rooted at each node. Existing OSPF/IS-IS style 
Shortest Path First (SPF) Dijkstra computations \cite{dijkstra} can be used to construct the tree.

\begin{figure}
\centering
\includegraphics[width=0.9\linewidth]{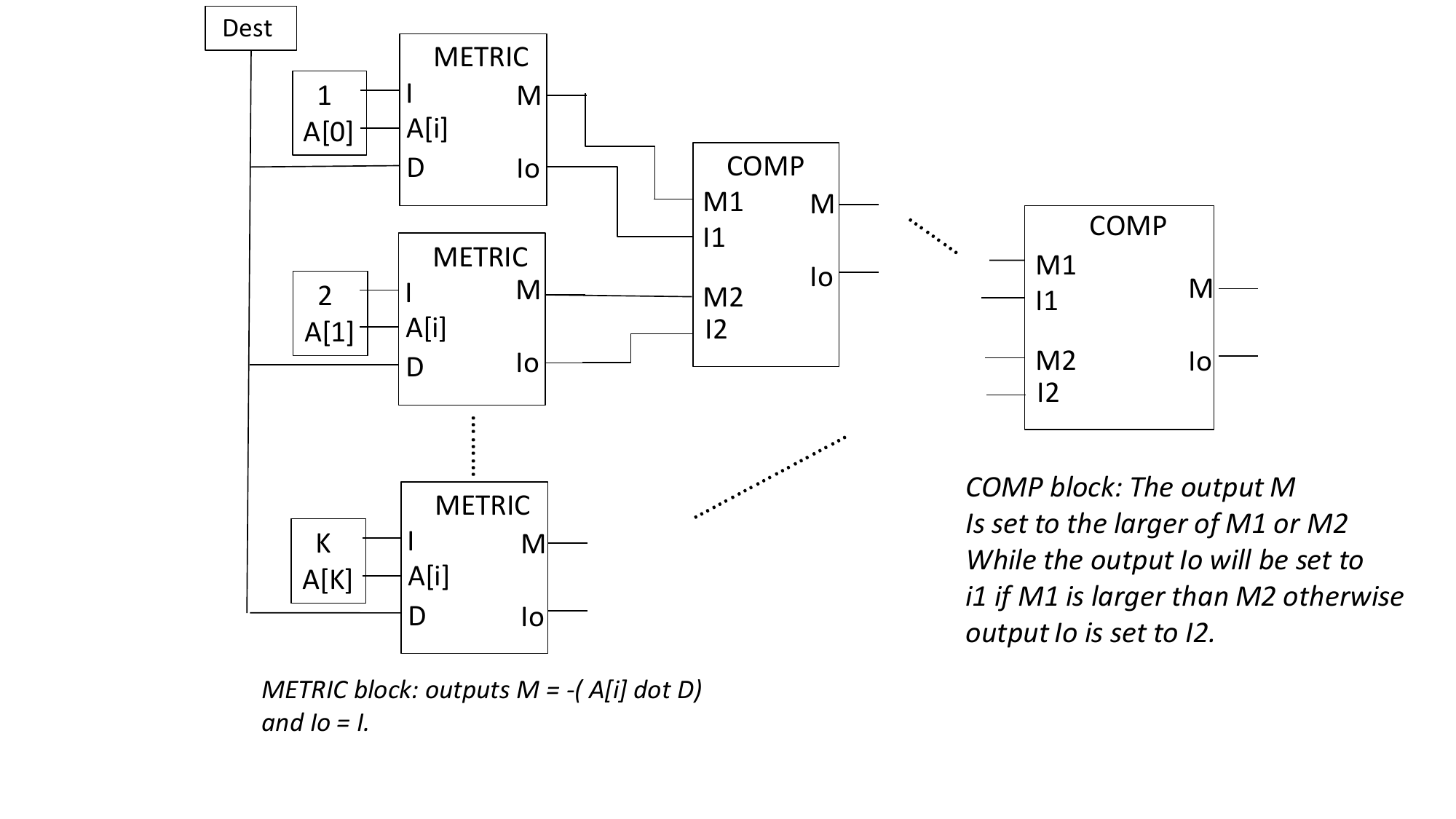}
\caption{OR(r) forwarding table comparator tree which can be implemented in hardware or parallel CPU processing in O(log(K)) clock cycles.}
\label{fig:forwarding-table}
\end{figure}

Forwarding then consists of finding the satellite I in this table which has the minimum $\hat{\mu}(\mbox{I}, D)$ metric and 
taking one hop towards it via the ISL indicated in the forwarding table.  See Figure \ref{fig:algorithm}.

Since the forwarding table is pre-computed this forwarding operation can be done in parallel in hardware by creating a tree of comparators 
of depth $\log_2(K)$ as shown in Figure \ref{fig:forwarding-table}.

The bottom of this tree compares in parallel all K satellite locations against the packet’s destination and each stage outputs the
metric to a second stage which compares those metrics two by two in parallel outputting the smaller to yet another stage. 
In this manner in $O(\log_2(K))$ clock cycles the smallest metric can be determined in an application specific IC or other 
hardware device.   Additional refinements could permit pipeline operations of the stages to further increase throughput.

For example 1000 satellites visible within our radius would mean about $O(10)$ cycles to find the closest entry for any given destination. 
This is quite reasonable for line rate forwarding hardware and implies that 10,000 satellites with billions of ground stations could 
function with a 1000 entry forwarding table (per satellite) and $O(10)$ cycle look-up times.

Satellites addresses/locations change according to a velocity vector that can either be advertised or known by other satellites. 
In between flood events (see section \ref{sec:flooding}), a satellite can apply incremental changes to the known satellites’ addresses 
according to their individual velocity vectors and the current 
time. In this manner the forwarding table’s addresses can be kept up to date at very high frequency during the intervals between floods. 
The granularity of these local recalculations of addresses depends on the speed of motion of the satellites and the time required to 
compute the new locations by the CPU within the satellite. Most likely this can be performed every 30 seconds for say 1000 satellites 
within its view/radius without much difficulty.

Using this addressing and forwarding method we can forward a packet one hop closer to the satellite which is closest to the given 
destination within the view/radius of the current satellite even when there are substantial failures (holes) in the mesh because we 
can route uphill as required within our radius $r$.

It should be clear that when the view/radius only extends to immediate neighbor satellites (i.e. $\mbox{OR}(1)$ ) the packet is forwarded 
in a greedy manner and the routing simply corresponds to geographic forwarding.  Such pure greedy algorithms only function correctly 
when there are no failures in the ISL mesh (all 4 links are up).

If the view/radius seen by the satellites is $\mbox{OR}(\infty)$ then we will be doing more or less link state forwarding directly to the satellite 
to which the ID in the destination is attached and failures in the mesh will not affect routing unless the network is partitioned. We don’t 
need any aggregation in this case however the forwarding tables will be immense and impractical.

The more interesting case is when we are forwarding to an intermediate ‘gateway’ which is closer to the destination but which 
does not have the destination ID attached while in the presence of failures in the ISL mesh. This constitutes the most difficult and 
most likely case of a very large network of satellites with some percentage of ISL links unavailable. This is where the limited 
radius/visibility floods are required. Figure \ref{fig:past-failure} depicts this case.

\subsection{Flooding}
\label{sec:flooding}

\orr uses floods of link state updates (LSUs) to distribute state information, similar to OSPF \cite{ospf}.   A LSU contains
information describing the state of the links at the originating satellite.

The topology database at each node will be different because the flooding radius $r$ should be considerably smaller than the 
number of satellites in the constellation.  LSUs originating from some node $C$ must occur on significant up/down link events but 
must also be periodic, with some period $V$.  A receiving node must age out LSUs from $C$ after $V$ has expired without a new 
update.   LSUs carry a TTL hop count to limit them to $r$ hops.  

LSUs are populated with basic data including adjacency information, and sent reliably to each adjacency.    Satellites will forward
LSUs to their neighbours until the TTL hop count has reached 0.  When a peer node receives a LSU it has to decide if it should forward it.  
If it already has a copy it just ignores it.  If the local copy is out of date it updates the local copy and forwards the LSU. 

In this manner each node will learn the state of the satellites and links within $r$ hops of itself. The smaller the 
value of $r$ the more frequently updates can be sent and therefore the more accurate the information within that 
radius will be. The object is to choose a value of $r$ which permits reasonable accuracy of routing within the radius 
while not creating too much load. We discuss the impact of $r$ later, but, in general, for a 12,000 node satellite 
constellation a radius containing several hundred satellites likely provides adequate failure handling.

\subsection{Path Computation}

The path computation from any given satellite is just an SPF tree from that satellite of depth $r$. This is a classic Dijkstra computation 
where every link is bidirectional and has a cost of 1. The cost used by Dijkstra is not related to the $\hat{\mu}$ metric.

Once an SPF tree has been computed to each of the satellites within radius $r$, the forwarding table is updated 
with the addresses of the satellites and the first hop ISL index toward them on the SPF tree.

Note that link state information which is more than $r$-hops away may still be present when a path computation is performed 
but will effectively be ignored due to the limit of a radius of $r$-hops in the computed SPF tree.

\subsection{Failure Behaviour}

\begin{figure}
\centering
\includegraphics[width=0.9\linewidth]{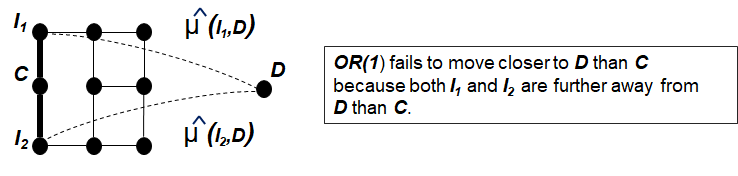}
\caption{OR(1) single link failure}
\label{fig:single-failure}
\end{figure}

Geographic routing i.e. $\mbox{OR}(1)$ scales to an infinite number of nodes but will fail when a node does not have all 4 ISLs up. 
Failed links/satellites create holes in the topology that cannot be bypassed by $\mbox{OR}(1)$.  A useful analogy is how 
simply moving progressively closer to some position on the earth may get you to the edge of an impasse (mountain or valley) 
which to detour around would require you to move further away from the destination temporarily. This failure is analogous 
to how multi-dimensional gradient following optimizations get stuck when a surface is non-convex. 
Figure \ref{fig:single-failure} shows this problem in a simple topology.

\begin{figure}
\centering
\includegraphics[width=0.9\linewidth]{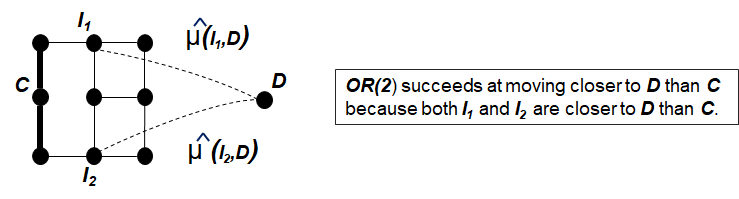}
\caption{OR(2) seeing past a link failure}
\label{fig:past-failure}
\end{figure}

The solution to this problem is to employ \orr where $r$ is sufficiently large to see around the impasse. 
Figure \ref{fig:past-failure} shows $\mbox{OR}(2)$ solving the problem of routing around the small hole caused by 
the single link failure in Figure \ref{fig:single-failure}.

At the small increased cost of learning the topology within just two hops $\mbox{OR}(2)$ is able to route around any number of holes 
bounded by 6 satellites in any sized constellation. 

This demonstrates an important property of \orr, its behavior is a function of the failure probabilities and the hole 
sizes but not the constellation size.

\begin{figure}
\centering
\includegraphics[width=0.9\linewidth]{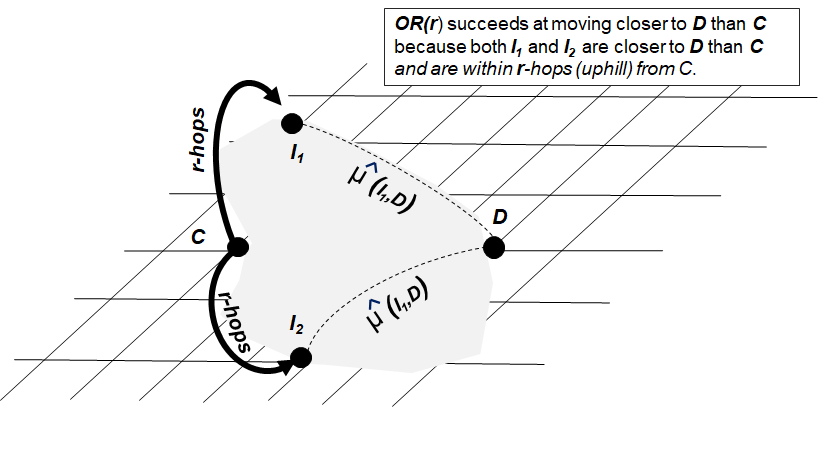}
\caption{OR(r) routing uphill}
\label{fig:routing-uphill}
\end{figure}

In general when there is a hole in the topology we need sufficiently large $r$ to be able to route uphill from the edge of that 
hole to a point that is now downhill to $D$ on the far side of the hole. This is shown in Figure \ref{fig:routing-uphill}.

Intuitively for a given link failure probability there should be a corresponding probability of holes of certain dimensions. 
An analytical solution to this problem has proven elusive and is for further research.

\subsection{Analysis}

Assume a bidirectional graph $G=(V,E)$ with nodal degrees $\le 4$ mapped onto the surface of a sphere such that:
\begin{itemize}
\item Vertex $v \in V$ has position in $R^3$ and dist() is the normal orthodromic distance between two vertices. 
\item $(v_i,v_j) \in E  \iff v_j$ is one of the closest by dist() vertices in $V$ to $v_i$.  
\item An edge metric $\mbox{Cost}(v_i,v_j)$ is defined for all edges and with value of $\{ \infty: \mbox{with probability} P;  
1 \mbox{ otherwise} \}$
\item The graph, $G,$ is connected.   Ignore the case of any $G$ that is disconnected as a result of link failures, as this $G$ is not
fully reachable.
\item Define $\mbox{DIST}(v_i, v_j)$ such that it breaks ties when two different node pairs have the same dist() 
between them.   There are many ways to accomplish this, such as
\[
\mbox{DIST}(v_i, v_j) = |E| dist(v_i, v_j) + |i-j|
\]

\end{itemize}

We can at some loss probability  $x$ compute a successful path from $v_S$ to $v_D$ using radius 
$r > 0$ hops:

\begin{algorithmic}[1]
\STATE $C \leftarrow S$
\WHILE{ True }
\IF{C=D}
\STATE stop(exact match) 
\ENDIF
\STATE $T \leftarrow \mbox{SpfTree}(\mbox{root}=C, \mbox{depth}=r,\mbox{metric}=\mbox{Cost()})$
\STATE $I \leftarrow \mbox{argmin} \mbox{ DIST}(I,D), I \in T$
\IF{I=C}
\STATE stop(can't get closer)
\ENDIF
\STATE Find unique path $\{ C, N, \ldots, I \}$ in tree $T$
\STATE $C \leftarrow N$
\ENDWHILE
\end{algorithmic}

{\bf Proposition 1:}  We show that the sequence of $I$'s that are chosen $I_1, I_2, \ldots$ must monotonicallys approach
D in DIST()

Steps 6, 7 and 11 are well understood problems \cite{dijkstra}, so no exposition is provided.

{\bf Proof by contradiction}.   Assume that $\mbox{DIST}(I_{i+1},D) > \mbox{DIST}(I_i,d).$   

\begin{itemize}
\item Let $C_i$ be the root of the SPF tree of depth $r$ which selected $I_i$ and $C_{i+1}$ be the 
root of the SPF tree which selected $I_{i+1}$ in step 7.   
\item $C_{i+1}$ is one hop closer to $I_i$ than $C_i$ (because of step 11) and $I_i$ is therefore in the SPF tree rooted 
at $Cc_{i+1}.$
\item $I_i$ is therefore in the SPF tree rooted at $C_{i+1}$
\item Step 7 cannot pick $I_{i+1}$ such that $\mbox{DIST}(I_{i+1},d)>\mbox{DIST}(I_i,d)$ since it cannot see $I_i$.
\end{itemize}

Therefore the set of $I$'s chosen must monotonically approach $d$ in DIST(), but not strictly.   We are left with the case
where $\mbox{DIST}(I_{i+1},D) = \mbox{DIST}(I_i,D).$

Since $\mbox{DIST}()$ requires that $I_{i+1}=I_i$ when $\mbox{DIST}(I_{i+1},D) = \mbox{DIST}(I_i,d)$ repeated values of 
DIST() can only occur with repeated values of $I$.  We need to show a bound on the number of times
a given I is chosen.

Let $C_1$ be the root of the SPF tree that first picks $I$.   The maximum number of times that $I$ can be chosen is the depth
of the SPF tree between $C_1$ and $I$.   This is because each choice $C_2, C_3, \ldots$ takes one step down the SPF tree towards 
$I$, which must cause the algorithm to stop when $C_n$ reaches $I$, or for a different $I$ closer to $d$ to be chosen somewhere 
along the $C_1, \ldots, I$ path.   Therefore there is a bound on the number of times a given $I$ can be chosen.

We can therefore state that \orr will progress in a greedy fashion closer and closer to $D$, and will not loop.  This does
not prove that it will reach $D$ in the presence of failures, except as $r \rightarrow |V|$.   We are primarily interested in
the minimum radius that can be proven to allow \orr to reach a destination at some probability with specific link failure
rates as this will be the most scalable $r$ for large constellations.

\section{Simulation Results}

\begin{figure}
\centering
\includegraphics[width=0.9\linewidth]{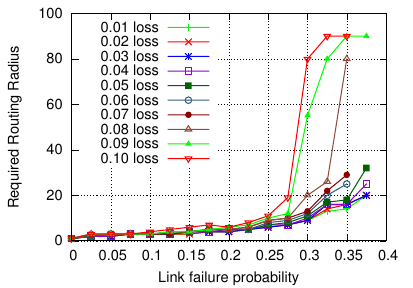}
\caption{Relationship between link failure probability and routing radius to produce various packet loss rates}
\label{fig:failure-prob}
\end{figure}

In order to understand failure behaviour at massive scales, we created a bespoke C++ simulator.   The simulator constructs a graph
corresponding to a Walker Delta satellite constellation with 4 ISLs per satellite as shown in Figure \ref{fig:algorithm}.    Each simulation
run generates random link failures, picks random ground based source and destinations, locates the corresponding overhead satellites, 
and runs the OR algorithm hop by hop across the graph from the source satellite to the destination satellite.   This was done
hundreds of thousands of times for different link failure probabilities, $0, 0.025, 0.050, ... , 0.375$ while increasing the 
routing radius from 1 to 100 hops.   

Figure \ref{fig:failure-prob} shows the required routing radius, $r$, as a function of the link failure probability for a given
packet loss rate.   The figure shows a relatively linear relationship until link failure probability exceeds 0.25.   At higher rates of
link failure, the routing radius grows more rapidly to the radius of the constellation.

The expected failure rates of ISL links are an open question.  If the ISL technology can maintain 75\% 
connectivity (over the regions of the globe of interest) then \orr can produce acceptably low packet loss rates between 
source destinations within those regions.

\section{Conclusions and Future Work}

For small satellite constellations of hundreds of satellites especially at higher orbits there is no reason why link-state protocols cannot be used.  However as the scale increases to thousands of satellites at lower orbits, link state protocols would normally be made hierarchical to limit the flood radius and path computation/forwarding table demands as the scale and event frequencies increase. It's not possible to have static hierarchy/gateways with moving satellites and so an alternative to limit the flood radius and aggregate addresses is required to achieve the desired scales. 

\orr routing creates a family of protocols which both limit the flood radius to $r$-hops while aggregating addresses based on proximity on the surface of a unit sphere. Simulations show that for very large constellations a modest radius of 10 hops can keep packet loss to under 1\% even with a 25\% link failure rate. \orr is therefore a practical approach to a fully distributed best effort protocol for satellite routing at very large scales.

\orr algorithms solve the two major challenges posed by very large dynamic satellite constellations; first they aggregate addresses naturally controlling forwarding table size; and second they keep the flood radius small to avoid heavy control plane messaging.

We are currently building a containerized \orr routing emulator to evaluate the protocol in detail.
Other areas for future research are development of an analytical model for the failure probability, and assessing the 
behaviour of \orr when the Dijkstra algorithm is using actual distances instead of hop count.


\begin{thebibliography}{00}
\bibitem{isl} A. Chaudhry, H. Yanikomeroglu, ``Laser Intersatellite Links in a Starlink Constellation: A Classification and Analysis,'' IEEE Vehicular Technology Magazine, Vol. 16, Issue 2, April 2021.
\bibitem{ospf} RFC 2328, ``OSPF Version 2,'', Internet Engineering Task Force, 1998.
\bibitem{isis} International Organization for Standardization, ``Intermediate system to Intermediate system intra-domain routeing information exchange protocol for use in conjunction with the protocol for providing the connectionless-mode Network Service,'' ISO/IEC 10589:2002, Second Edition, 2002.
\bibitem{coverage} Y. Gong, S. Zhang, X. Peng, ``Quick coverage analysis of mega Walker constellation based on 2D map,'' Acta Astronautica, Volume 188, 2021, pages 99-109.
\bibitem{aodv} N. J. H. Marcano, J. G. F. Nørby and R. H. Jacobsen, ``On Ad hoc On-Demand Distance Vector Routing in Low Earth 
Orbit Nanosatellite Constellations," 2020 IEEE 91st Vehicular Technology Conference (VTC2020-Spring), 2020, 
pp. 1-6, doi: 10.1109/VTC2020-Spring48590.2020.9128736.
\bibitem{virtual-node}E. Ekici, I. F. Akyildiz and M. D. Bender, ``Data-gram routing algorithm for LEO satellite networks,” 
The 19th Annual Joint Conf. of the IEEE Comp. and Comm. Societies, pp. 2: 500-508, 2000.
\bibitem{demand-island} O. Markovitz and M. Segal, ``Advanced Routing Algorithms for Low Orbit Satellite Constellations," ICC 2021 - IEEE International Conference on Communications, 2021, pp. 1-6, doi: 10.1109/ICC42927.2021.9500740.
\bibitem{link-state} L. Zhang et al., ``A Routing Algorithm Based on Link State Information for LEO Satellite Networks," 2020 IEEE Globecom Workshops (GC Wkshps, 2020, pp. 1-6, doi: 10.1109/GCWkshps50303.2020.9367496.
\bibitem{scale1} M Goyal, M Soperi, E Baccelli, G. Choudhury, A. Shaikh, ``Improving Convergence Speed and Scalability in OSPF: A Survey,'' Communications Surveys and Tutorials, IEEE Communications Society, 2012, 14(2), pp.443-463.
\bibitem{scale2} M Karakus, A Durresi, ``A survey: Control plane scalability issues and approaches in Software Defined Networking,'' Computer Networks, Volume 112, 2017, pp279-293.
\bibitem{dijkstra} E Dijkstra, ``A note on two problems in connexion with graphs,'' Numerishce Mathematik,  1:269-271.
\end{thebibliography}
\end{document}